\def \cm{~\rm{cm}}
\def \s{~\rm{s}}
\def \km{~\rm{km}}

\def \K{~\rm{K}}
\def \g{~\rm{g}}

\def \AU{~\rm{AU}}
\def \erg{~\rm{erg}}

\def \yr{~\rm{yr}}

\def \kpc{~\rm{kpc}}
\documentclass[12pt,preprint]{aastex}
\usepackage{natbib}

\shorttitle{Pairs of bubbles}
\shortauthors{Soker}
\slugcomment{Draft version of \today}

\begin{document}


\title{BUBBLES IN PLANETARY NEBULAE AND CLUSTERS OF GALAXIES:
JET PROPERTIES}

\author{Noam Soker\altaffilmark{1}}

\altaffiltext{1}{ Department of Physics, Oranim, and
Dept. of Physics, Technion, Haifa 32000, Israel;
soker@physics.technion.ac.il.}

\begin{abstract}

I derive constraints on jet properties for inflating pairs
of bubbles in planetary nebulae and clusters of galaxies.
This work is motivated by the similarity in morphology and some
non-dimensional quantities between X-ray-deficient bubbles in clusters
of galaxies and the optical-deficient bubbles in planetary nebulae,
which was pointed out in an earlier work.
In the present paper I find that for inflating fat bubbles, the
opening angle of the jets must be large, i.e., the half opening
angle measured from the symemtry axis of the jets should typically be
$\alpha \gtrsim 40 ^\circ$. 
For such wide-opening angle jets, a collimated fast wind (CFW)
is a more appropriate term.
Narrow jets will form elongated lobes rather than fat bubbles.
I emphasize the need to include jets with large opening angle,
i.e., $\alpha \simeq 30-70^\circ$, in simulating bubble inflation
in both planetary nebulae and (cooling flow) clusters of galaxies.

\end{abstract}


\section{INTRODUCTION} \label{sec:intro}

{\it Chandra} X-ray observations of clusters of galaxies reveal the
presence of X-ray-deficient bubbles in the inner regions of many
(cooling flow) clusters, e.g.,
Hydra A (McNamara et al.\ 2000), Perseus (Fabian et al.\ 2000, 2003),
A 2597 (McNamara et al.\ 2001), RBS797 (Schindler et al.\ 2001),
Abell~4059 (Heinz et al.\ 2002), and Abell~2052, (Blanton et al. 2003).
A nice pair of bubbles is seen in the poor cluster HCG 62
(Vrtilek et al.\ 2002).
X-ray-deficient pairs of bubbles, although less prominent, exist
also in elliptical galaxies; e.g., in M84 (Finoguenov \& Jones 2001)
and NGC 4125 and NGC 4552 (White \& Davis 2003).
The low X-ray emissivity implies low density inside the bubbles, while
the absence of evidence of strong shocks suggests that the bubbles are 
expanding and moving at subsonic or mildly transonic velocities
(Fabian et al.\ 2000, 2003; McNamara et al.\ 2000;
Blanton et al.\ 2001).
In paper-I (Soker 2003) I pointed out an interesting and not trivial
similarity in the morphology and some non-dimensional quantities
between pairs of X-ray-deficient bubbles in clusters of galaxies and
pairs of optical-deficient bubbles in planetary nebulae (PNs).
Examples of PNs with nice pairs of bubbles are
the Owl nebula (NGC 3587; PN G148.4+57.0: 
Guerrero et al.\ 2003),  Cn 3-1 (VV 171; PN G038.2+12.0;
Sahai 2000), and Hu 2-1 (PN G051.4+09.6), which possesses
two prominent pairs of bubbles, one pair closer to the center and the
other farther out, with inclination between the two symmetry axes
(Miranda et al.\ 2001).
Many other PNs also possess pairs of low emissivity bubbles,
although less prominent, e.g., NGC 2242 (PN G170.3+15.8:
Manchado et al.\ 1996), and a pair of bubbles within bipolar
lobes in M 2-46 (PN G024.8-02.7: Manchado et al.\ 1996). 
More details of the morphological similarities and the similarity in
some non-dimensional variables between pair of bubbles in these two
classes of objects is given in paper I (see Table 1 there).

The similarity in morphology and in some relevant non-dimensional variables
led me to postulate in paper I a similar formation mechanism.
It is commonly accepted that bubbles in clusters are blown by AGN jets 
(e.g., Brighenti \& Mathews 2002; Br\"uggen 2003; Br\"uggen et al.\ 2002;
Fabian et al.\ 2002; Nulsen et al.\ 2002; Quilis, Bower, \& Balogh 2001;
Reynolds, Heinz, \& Begelman 2001; Soker, Blanton, \& Sarazin 2002).
This postulate, then, was used to strengthen models for PN shaping by jets
(or collimated fast winds: CFW).
In particular, the presence of dense material in the equatorial plane
observed in the two classes of bubbles constrains the jets and CFW
activity in PNs to occur while the AGB star still blows its dense wind,
or very shortly after.
I argued there  that only a stellar companion can account for such
fast jets (or CFW).
Very recent {\it Chandra} observations hint at more similarities
between the jets in PNs and AGN jets.
Kastner et al. (2003) point out that the blobby X-ray appearance of the
jets in the symbiotic bipolar nebula Mz 3, probably to become a PN,
is similar somewhat to the blobby X-ray nature of some AGN jets, e.g.,
Centaurus A (Hardcastle et al. 2003).
The similarity is not only of fat bubbles in cluster and bipolar PNs.
The general X-ray image of the cooling flow cluster Cygnus A
(Smith et al.\ 2002; see their fig. 2) resembles that of the optical
image of many elliptical PNs, e.g., NGC 6826 (Balick 1987;
Balick et al.\ 1998).
Cygnus A hot spots, i.e., the heads of the two jets, resemble the
structure of pairs of ansae$-$two opposite dense blobs along the
major axis of some elliptical PNs$-$suggesting that the ansae are
formed by jets (see for example the ansae of the PN NGC 7009;
Balick 1987; Balick et al.\ 1998;  Sabbadin et al.\ 2003).

These similarities motivate me to consider a unified scheme for the
formation of bubbles in clusters and PNs, based on the inflation
of pairs of bubbles by oppositely ejected jets.
Although commonly accepted by people studying clusters, the jet-inflated
bubble model is still in dispute among people studying PNs.
In particular, it is not widely accepted that jets which shape bipolar
PNs$-$those with two large lobes and an equatorial waist between
them$-$are blown by a binary companion.
As argued in several papers (Paper I and references therein), more
and more observations, including the similarity in morphology
discussed above, point to jets blown by a binary companion in
bipolar PNs (but not in all PNs).

In the present paper I study in more detail the conditions under
which, more or less, spherical bubbles are formed, as opposed to
the case where a propagating jet inflates a long narrow region.
A `fat' low-density inflated region, more or less spherical but
not necessarily exactly so, is termed in the present paper a fat bubble
(e.g., as in the Owl nebula), while an elongated low density inflated
region is termed lobe, e.g., as in OH 231.8+4.2 (Bujarrabal et al.\ 2002),
and He2-115 (Sahai \& Trauger 1998).

\section{JET PROPAGATION IN PLANETARY NEBULAE}
\label{pnjet}

The flow structure is of a jet flowing into the dense circumstellar
material, which is the expanding AGB wind.
The speed of the material inside the jet is $v_j$, while the jets'
head propagates at speed $v_h<v_j$. The slow wind speed is
$v_s<v_h$. 
The aim here is to examine the conditions for the formation of
fat bubbles.
The condition is that the expansion speed of the bubble surface
relative to the slow wind $v_b$ be faster than the propagation
speed of the jet's head relative to the slow wind
\begin{equation}
v_{hs} \equiv v_h-v_s < v_b \qquad {\rm for~spherical~bubble~formation.}
\end{equation}
If, on the other hand, $v_{hs} > v_b$, the expanding bubble formed
by the jet lags behind the jet's head, and an elongated lobe,
rather than a fat bubble, is formed (Soker 2002; Lee \& Sahai 2003).
The condition in equation (1) implies that the jet is slow-propagating,
with its velocity relative to the slow wind speed given by
momentum conservation, which here reads (see eq. 1 in Soker 2002)
\begin{equation} 
v_{hs} \simeq  
\left( \frac{\dot M_j v_j }{\beta \dot M_s v_s} \right)^{1/2} v_s, 
\qquad {\rm for} \qquad \rho_s \gg \rho_j.   
\end{equation}
where $\beta$ is defined such that the jet expands into
a solid angle of $\Omega_j = 4 \pi \beta$, and $\rho_s(r)$ and
$\rho_j(r)$ are the density of the slow wind and jet at a distance
$r$ from the center, respectively.
$\dot M_s$ and $\dot M_j$ are the mass loss rates into the
slow wind and one jet, respectively; both are constant,
and defined positively.

The jet deposits kinetic energy in a power of $\sim 0.5 \dot M_j v_j^2$
into the bubble.
The radius of the bubble as a function of time $t$, can be
approximated by using the results of Castor, McCray, \& Weaver (1975),
\begin{equation}
R_b =  0.76 \left( \frac{0.5 \dot M_j v_j^2}{\rho_s} \right)^{1/5} t^{3/5},
\end{equation}
where $\rho_s=\dot M_s/(4 \pi v_s r^2)$.
The expansion velocity of the bubble's surface relative to the slow wind is
\begin{eqnarray}
v_b = 0.6 R_b/t =  18 
\left( \frac {v_j}{400 \km \s^{-1}} \right)^{2/5} 
\left( \frac {\dot M_j} {0.01 \dot M_s} \right) ^{1/5}
\left( \frac {v_s} {10 \km \s^{-1}} \right)^{1/5} 
\nonumber \\ \times
\left( \frac {z} {10^{17} \cm} \right) ^{2/5} 
\left( \frac {t} {1000 \yr } \right)^{-2/5} \km \s^{-1} ,  
\end{eqnarray}
where the expression for $\rho_s$ was used, and $z$ is the coordinate
along the propagation direction of the jet, with the origin
at the origin of the jet.
With the parameters used in the last equation, equation (2) gives
$v_h \sim 10 [1+2 (\beta/0.1)^{-1/2}] \km \s^{-1} \sim 30 \km \s^{-1}$,
where $\beta=0.1$ is for a jet with a half opening angle
(i.e., measured from its symmetry axis) of $\alpha = 37 ^\circ$.
The distance of the jet's head from the source at time $t$
is given by
\begin{equation} 
z_h \sim 10^{17} \left[ \frac{1}{3} + \frac{2}{3}
\left( \frac{\beta}{0.1} \right)^{-1/2}
\right] \left( \frac{t}{1000 \yr} \right) \cm .
\end{equation}
This is the reason for the scaling of $z$ in equation (4).
Note that what matters is the ratio $z/t$, which changes slowly,
if at all, with time.
Substituting $v_{hs}$ from equation (2), and $v_b$ from equation (4)
in condition (1) for inflating a fat bubble, gives
the condition of spherical bubble formation in the form
\begin{eqnarray} 
1 \lesssim  \frac{v_b}{v_{hs}} = 0.9  
\left( \frac {v_j}{400 \km \s^{-1}} \right)^{-1/10} 
\left( \frac {\dot M_j} {0.01 \dot M_s} \right) ^{-3/10}
\left( \frac{\beta}{0.1} \right) ^{1/2}
\left( \frac {v_s} {10 \km \s^{-1}} \right)^{-3/10}
\nonumber \\ \times
\left( \frac {z} {10^{17} \cm} \right) ^{2/5} 
\left( \frac {t} {1000 \yr } \right)^{-2/5} \km \s^{-1}.  
\end{eqnarray}
From equation (5) the ratio $z/t$ does not change much in
different relevant cases. Also, the slow wind speed is
$v_s \sim 10 \km \s^{-1}$ for all relevant AGB stars.
The condition for fat-bubble inflation becomes mainly a condition
on the jet's properties
\begin{eqnarray} 
\beta \gtrsim 0.12 
\left( \frac {v_j}{400 \km \s^{-1}} \right)^{1/5}
\left( \frac {\dot M_j} {0.01 \dot M_s} \right) ^{3/5}.
\end{eqnarray}
Namely, more or less spherical fat bubbles will be inflated by non-well
collimated jets.
For these jets, a more appropriate term is CFW (collimated fast wind).

For small angles $\beta \simeq \alpha^2/4$, and the last
condition can be written for the half opening angle of the jet
\begin{equation} 
\alpha \gtrsim 40 ^\circ 
\left( \frac {v_j}{400 \km \s^{-1}} \right)^{1/10}
\left( \frac {\dot M_j} {0.01 \dot M_s} \right) ^{3/10}.
\end{equation}
The scaling of the jet (or CFW) properties used here is appropriate for
a jet blown by a main sequence companion, because the jet's speed
is of the order of the escape velocity.
The mass loss rate into the jets$-$two oppositely blown jets$-$is
quite high, e.g., a companion accreting a fraction of $0.2$ of
the AGB wind, and blows a fraction of $0.1$ of the accreted mass
into the two jets.
In this case, the last condition implies that the jets, or more
appropriately a CFW, will be non-well collimated.
This is reasonable for a close main sequence companion, since
the accretion disk will not be extended.
The main sequence companion can resides at larger distances, but
not too large $a \lesssim 20 \AU$, otherwise the jet will be weak
and the bubble will not be inflated much.

At the other extreme the jets may be well collimated, and
their mass deposition rate very low
$\dot M_j \ll 0.01 \dot M_s$.
For the bubbles to be fat, the jet's speed must be high.
For example, for  $\dot M_j = 10^{-4} \dot M_s$ and
$v_j = 5000 \km \s^{-1}$, condition (8) reads
$\alpha \gtrsim 10^\circ$.
These parameters are appropriate for a widely separated,
$a \sim 50 \AU$, white dwarf companion.

For a large pair of bubbles to be observed in PNs there are two other
conditions, in addition to the condition derived here.
First, the jet should be strong enough to inflate a large bubble that
expands fast enough, i.e., faster than the expansion speed of the
slow wind,
\begin{equation} 
v_{hs} \gtrsim v_s .
\end{equation} 
The condition on the jet's properties implied by this inequality can be
derived from equation (4).
Second, the jet's material should not cool fast.
This condition is that the bubble starts to expand close to the
center, $z \lesssim 10^{17} \cm$.
The location where the bubble starts to expand is given by equation
(6) in Soker (2002).
Basically, keeping the other parameters as used in this section,
it constrains the jet's speed to 
\begin{equation} 
v_j \gtrsim 150 \km \s^{-1}.
\end{equation} 

Finally, I note that the conditions on the jet's properties given
in equations (8) (or 6), (9) and (10), were derived for the flow occurring
before ionization and the spherical fast wind from the central
post-AGB star start. These are likely to erode the bubble; in particular,
after ionization the thermal pressure of the slow wind material can't be
neglected any longer.  Therefore, the real constraints on jets to inflate
observable fat bubbles in PNs are somewhat stronger than those given
by these equations.

\section{JET PROPAGATION IN CLUSTERS}
\label{cljet}

There are several significant differences between bubble inflation
in PNs and clusters.
(1) In clusters the thermal pressure of the ambient
gas is non-negligible.
(2) In clusters the ambient medium does not flow outward.
(3) The inflating jets in clusters may be relativistic, and the magnetic
pressure inside the bubble can be large.
(4) In clusters the bubbles can be observed as they form, unlike in PNs,
where they are observed long after the jets have ceased
(old bubbles may be observed in clusters$-$termed ghost-bubbles$-$as in
the Perseus cluster; Fabian et al. 2000). 
However, as argued in paper I, these don't prevent a similar
bubble-formation mechanism in PNs and clusters.

Following Soker et al. (2002) I neglect the ambient pressure
and use the results of Castor et al.\ (1975), as in the previous section.
The expansion velocity of the bubble is given by equation (20) of
Soker et al.\ (2002), which is written here as 
\begin{equation}
v_b (\tau_I ) \simeq 720  
\left(\frac {\tau}{10^7 \, {\rm yr}} \right)^{-2/5} 
\left(\frac {\dot E_j}{10^{45} \, {\rm erg} \s^{-1} } \right)^{1/5} 
\left(\frac {\rho_c}{10^{-25} \, {\rm g} \, {\rm cm}^{-3}} \right)^{-1/5}
\, {\rm km} \, {\rm s}^{-1} .
\end{equation}
For the ambient density a crude approximation for $r > 10 \kpc$
in A 2052 is (Blanton et al. 2001; Soker et al. 2002, their eq. 16)
\begin{equation}
\rho_c (r) \approx 10^{-25}
\left( \frac{r}{10 \kpc} \right)^{-1} 
\g \cm^{-3}.
\end{equation}
Because of the presence of bubbles, it is impossible to know the density
prior to bubble inflation in A 2052, so this approximation
is more accurate for $r>30 \kpc$ and very crude for smaller radii.
By neglecting the magnetic pressure inside the jet and relativistic effects,
hence $\dot E_j = \dot M_j v_j^2/2$, we can take an equation similar to
equation (2) in the previous section, and write for speed of
the jet's head through the cluster medium (e.g., Krause 2003) 
\begin{eqnarray}
v_h \simeq \left(\frac {\dot E_j}{2 \pi \beta z^2 v_j \rho_c} \right)^{1/2}
= 1300  
\left( \frac {\dot E_j} {10^{45} \erg \s^{-1}} \right)^{1/2}
\left( \frac {v_j}{10^4 \km \s^{-1}} \right)^{-1/2} 
\nonumber \\ \times
\left( \frac{\beta}{0.1} \right) ^{-1/2}
\left(\frac {\rho_c}{10^{-25} \g  cm^{-3}} \right)^{-1/2}
\left( \frac {z} {10 \kpc} \right) ^{-1} \km \s^{-1}
\end{eqnarray}
where the symbols have their meaning as in section 2.
For the parameters used here, at $z \sim 10 \kpc$ the cluster density
to jet density ratio is $\sim 100$. 
We scaled the jet velocity to the non-relativistic regime, as
Omma et al. (2003) do. 
The velocities (which depend on the time and location) $v_b$
and $v_h$ given above, are mildly supersonic, or even subsonic.
This means the ambient pressure cannot be neglected for late time.
Still, there is justification for neglecting pressure
(Soker et al. 2002), in particular in the approximate
condition derived below.
An accurate expression cannot be derived because of the
poor knowledge of the exact density profile in the very inner regions
of the cluster.
For the scaling used above, the jet reaches a distance of
$r=10 \kpc$ in $\sim 1.4 \times 10^7 \yr$. Therefore, the
time and distance scaling used above are compatible.
Inserting equations (11) and (13) in the condition to inflate
a fat bubble $v_b \gtrsim v_h$ (eq. 1), I derive a condition on the half
opening angle of the jet (similar to that used in deriving eq. 8),
\begin{eqnarray}
\alpha \gtrsim 65 ^\circ 
\left( \frac {\dot E_j} {10^{45} \erg \s^{-1}} \right)^{3/10}
\left( \frac {v_j}{10^4 \km \s^{-1}} \right)^{-1/2} 
\left(\frac {\rho_c}{10^{-25} \g  cm^{-3}} \right)^{-3/10}
\left( \frac {z} {10 \kpc} \right) ^{-1} 
\left(\frac {\tau}{10^7 \rm yr} \right)^{2/5} .
\end{eqnarray}

Because several factors were neglected, as mentioned earlier,
the last condition is a crude one. Still, a strong constraint
can be deduced from it, namely in order to inflate a fat bubble
in cooling flow clusters the CFW should be wide open.
Even for $\dot E_j \simeq 10^{44} \erg \s^{-1}$ and
$v_j \simeq 3 \times 10^4 \km \s^{-1}$, equation (14) gives
$\alpha \gtrsim 20 ^\circ$.
The jets which inflated the bubbles in the cluster A2052 were
probably such jets.
(Weak relativistic jets can be narrow though.)
This may point to a significant difference between the AGN activity in
normal large elliptical galaxies, and the AGN activity of the
cD galaxies in cooling flow clusters. 
In any case, keeping in mind the uncertainties in the derivations of
the last two sections, the conditions on the opening angle of the
CFW (jets) in PNs and clusters to inflate fat bubbles are similar.
This further strengthen the idea of a unified process.

\section{COMPARISON WITH NUMERICAL SIMULATIONS} \label{sec:compare}

Most numerical simulations of jets aim at a specific environment.
I could not find in the literature simulations of conical, rather than
cylindrical, jets propagating into a denser medium in the
parameter-space relevant to the present cases.
This is the main motivation for the order-of-magnitude estimates
presented in this paper: To encourage simulations with
wide-angle jets.

The most relevant simulations in the literature are in a very recent
paper by Omma et al.\ (2003), who conduct numerical simulations of
non-relativistic jets propagating into a cooling flow cluster
environment.
They simulate two cases where they inject cylindrical jets,
one case with a jet radius of $r_{\rm jet}=2 \kpc$,
and the other with $r_{\rm jet}=3 \kpc$.
Close to the source, where the distance along he jet axis is
not much larger than the jet's diameter $z \lesssim 2 r_{\rm jet}$,
they obtain an inflated fat-bubble.
As the cylindrical jet expands to larger distances,
$z \gtrsim 2 r_{\rm jet}$, the bubble loses its general spherical
structure, and it is detached from the center.
Such simulations can't reproduce fat bubbles as in Perseus or A2052,
unless the radius of the simulated jets at injection is very large.
They do reproduce structures similar to the detached (from the center)
bubbles in M 87 (Virgo A; See fig. 15 of Omma et al.\ 2003) and Hydra A.
The wider jet in the simulations of Omma et al.\ (2003) forms
fatter bubbles and proceeds at a lower velocity.
This implies that at a specific location in the simulated
cluster environment,  when the jet radius is larger than some
critical radius, a bubble is formed rather than a narrow cocoon.
However, eventually the jet proceeds forward faster than the bubble
expands (until it slows down when injection ends).
To maintain a bubble attached to the center, or close to it,
the jet radius should increase, i.e., a conical jet, as was shown
in the previous section of the present paper.

A more quantitative comparison is possible.
The ambient density, as they take from David et al.\ (2001),
is $\rho_c \simeq 10^{-25} \g  cm^{-3}$, and their
injection power and velocity are
$\dot E_j = 6 \times 10^{43} \erg \s^{-1}$ and $V_j=10^4 \km \s^{-1}$,
respectively. 
Substituting these values in equation (14) gives
$\alpha \gtrsim 28 ^\circ$, for the same scaling of time and distance
as in equation (14).
For the  $r_{\rm jet}=2 \kpc$ jet this implies $z \simeq 4 \kpc$,
while for $r_{\rm jet}=3 \kpc$ this implies $z \simeq 6 \kpc$.
This value of $r_{\rm jet}/z \sim 0.5$ explains why the bubbles
in Omma et al.\ (2003) simulations lose their `fatty' structure
when the jet reaches a distance of about the jet diameter. 
Although it is hard to compare the cylindrical-jet simulations
of Omma et al.\ (2003) with the present paper, the above discussion
suggests that there is a solid ground for the order of magnitude
analysis conducted here.

There are less relevant simulations, which non the less shed
some light on the present analysis; for these, only qualitative
comparison is meaningful.
Krause \& Camenzind (2003b) find that as they inject the jet
a spherical bubble is formed around the jet.
Later the bubble becomes elongated, first to an elliptical shape,
then to a narrow extension along the jet.  This is compatible
with the results of Omma et al.\ (2003) cited above.
This is along the claim of the present paper that a wide-angle
jet will form a long-lasting large spherical (fat) bubble.
Loken et al.\, (1995) simulate the formation of wide-angle tailed
radio sources, which are though to form when a radio galaxy which 
blows two jets moves through the intracluster medium. 
This is quantitatively similar to precessing jets, in that the jets 
continuously encounters fresh ambient medium.
Their three-dimensional simulations clearly show how jets are 
disrupted, and large bubbles are formed. 
These simulations support the claims that wide-angle jets
or precessing jets can be disrupted more easily and form
fat bubbles. 
Lee et al.\ (2001) conduct isothermal simulations of jets and winds
from young stellar objects.
Despite that isothermal flows are less favorable to inflate
bubbles than the cases studied here where radiative
cooling time is long, Lee et al.\ (2001) find that the overall width
of the jet-driven shell is smaller than that of the wide-angle
wind-driven shell.
This supports the notion that wider flows form fatter bubbles
(shells in their case).

Finally, it should be noted that recollimation of wide-angle jets,
such that the opening angle $\alpha$ significantly decreases, 
will not change much the conclusions.
To recollimate a continuous jet the ambient medium must impart
transverse momentum to the jet's material.
For a wide-angle jet, the transverse speed is not much below the
jet's speed, a factor of $\sim \sin \alpha$.
Therefore, the ambient medium and the jet's material will
go through strong shocks, heat-up, and form hot low density
bubbles, as assumed in the previous sections.

\section{DISCUSSION AND SUMMARY} \label{sec:conclusion}

Paper I pointed out the similarities in morphology and some
non-dimensional quantities between X-ray-deficient bubbles in
clusters of galaxies and the optical-deficient bubbles in PNs.
The comparison here, as in paper I, is for pairs of fat,
almost spherical, bubbles in these two classes of objects.
The main goal of the first paper was to point out these
similarities; from that I argue that pairs of fat bubbles in PNs are
formed by jets, most probably blown by a companion.
In the present paper the main goal is to examine some similarities
between the jets in these two classes of systems, assuming indeed that
jets, or CFW (for collimated fast wind) blow the fat bubbles.
The main conclusion is that for such bubbles to be formed, in
both classes of systems the opening angle of the
jets must be large.
In equation (8) for PNs, and equation (14) for clusters,
the values for the half opening angle (measured from the symmetry axis)
for blowing fat bubbles are given.
For such wide-opening angle jets, a CFW is the more appropriate term.
Narrow jets will form elongated lobes rather than fat bubbles
(Krause 2003; Krause \& Camenzind 2003a);  
some examples of these are given in section 1.
Presently, most simulations (e.g, Krause 2003;  Krause \& Camenzind 2003a,
and references in these papers) simulate narrow jets, and examine
the influence of the jet to ambient density ratio.
The present study suggests that the opening angle is a crucial parameter.

Beside the similarities studied in paper I and here, other similar
processes may exist. An example is heat conduction from the hot
bubbles to the cooler environment.
In the past this process was studied in more detail for PNs
(Soker 1994; Zhekov \& Myasnikov 2000; Soker \& Kastner 2003).
This process, among other things, was claimed to be able to heat
somewhat the cooler ($\sim 10^ 4\K$) environment.
The heating will not be isotropic because it is regulated by
magnetic fields (Soker 1994; Zhekov \& Myasnikov 2000).
With the search for heating sources of the radiatively cooling ICM 
in cooling flow clusters, heat conduction from the hot bubble to the
cooler ($\sim 1-7 \times 10^7 \K$) ICM brought this process to
the attention of the cooling flow community. 
The hot spots found around the X-ray depression region in the
cooling flow cluster in MKW3s (Mazzotta et al.\ 2002) may have been
heated by such heat conduction, as suggested by some of these authors
during the Cooling Flow meeting held in Charlottesville, Virginia,
in June 2003.

I summarize by emphasizing again  the need to include jets with
large opening angles, i.e., half opening angles of
$\alpha \simeq 30-70^\circ$, in simulating both PNs and
cooling flow clusters (and galaxies).
Such jets are more appropriately called CFW.

\acknowledgements
This research was supported in part by the Israel Science Foundation.


\begin{references}


\reference{} Balick, B. 1987, AJ, 94, 671 

\reference{} Balick, B., Alexander, J., Hajian, A., R., Terzian, Y., 
      Perinotto, M., \& Patriarchi, P. 1998, AJ, 116

\reference{} Blanton, E. L., Sarazin, C. L., McNamara, B. R., \&
Wise, M. W. 2001, ApJ, 558, L15  

\reference{} Blanton, E. L., Sarazin, C. L., \& McNamara, B. R.
   2003, ApJ, 585, 227

\reference{} Brighenti, F., \& Mathews, W. G. 2002, ApJ, 574, L11

\reference{} Br\"uggen, M. 2003, ApJ, 592, 839 

\reference{} Br\"uggen, M., Kaiser, C. R., Churazov, E., \& Ensslin, T. A.
        2002, MNRAS, 331, 545 
 
\reference{} Bujarrabal, V., Alcolea, J., S\'anchez Contreras, C.,
     \& Sahai, R. 2002, A\&A, 389, 271
  
\reference{} Castor, J., McCray, R., \& Weaver, R. 1975, ApJ, 2000, L107
 
\reference{} David, L.\ P.\, Nulsen, P.\ E.\ J.\, McNamara, B.\ R.\, 
    Forman, W.\, Jones, C.\, Ponman, T.\, Robertson, B.\, \& Wise, M.\
    2001, ApJ, 557, 546

\reference{} Fabian, A. C., Celotti, A., Blundell, K. M., Kassim, N. E., \&
  Perley, R. A. 2002, MNRAS, 331, 369

\reference{} Fabian, A. C., Sanders, J. S., Crawford, C. S., 
   Conselice, C. J., Gallagher III, J. S., \& Wyse, R. F. G. 2003,
   MNRAS, in press (astro-ph/0306039)
  
\reference{} Fabian, A. C., et al.\ 2000, MNRAS, 318, L65 

\reference{} Finoguenov, A., \& Jones, C. 2001, ApJ, 547, L107
  
\reference{} Guerrero, M. A., Chu, Y.-H., Manchado, A., Kwitter, K. B.
       2003, AJ, 125, 3213

\reference{} Hardcastle, M. J., Worrall, D. M., Kraft, R. P., Forman, W. R.,
   Jones, C., \& Murray, S. S. 2003, ApJ, in press (astro-ph/0304443) 

\reference{} Heinz, S., Choi, Y.-Y., Reynolds, C. S., \& Begelman, M. C. 2002,
ApJ, 569, L79

\reference{} Kastner, J. H.,  Balick, B., Blackman, E. G., Frank, A.,
Soker, N., Vrtilek, S. D., \& Li, J.  2003, ApJ, 591, L37
 
\reference{} Krause, M. 2003, A\&A, 398, 113 

\reference{} Krause, M. \& Camenzind, M. 2003a,
New Astron. Rev., proceedings of the conference in Bologna: 
"The Physics of Relativistic Jets in the CHANDRA and XMM Era",
(astro-ph/0301289) 

\reference{} Krause, M. \& Camenzind, M. 2003b, (astro-ph/0307152)

\reference{} Lee, C.-F., \& Sahai, R. 2003, ApJ, 586, 319

\reference{} Lee, C.-F.\, Stone, J.\ M.\, Ostriker, E.\ C.\, 
     \& Mundy, L.\ G.\ 2001, ApJ, 557, 429  

\reference{} Loken, C.\, Roettiger, K.\, Burns, J.\ O.\, \&
  Norman, M.\ 1995, ApJ, 445, 80

\reference{} Manchado, A., Guerrero, M., Stanghellini, L.,
\& Serra-Ricart, M. 1996, The IAC Morphological Catalog of Northern
Galactic Planetary Nebulae (Tenerife: IAC).

\reference{} Mazzotta, P., Kaastra, J. S., Paerels, F. B., 
   Ferrigno, C., Colafrancesco, S., Mewe, R., \& Forman, W. R. 
       2002, ApJ, 567, L37

\reference{} McNamara, B. R., et al.\ 2000, ApJ, 534, L135 

\reference{} McNamara, B. R., et al.\ 2001, ApJ, 562, L149 

\reference{} Miranda, L. F., Torrelles, J. M., Guerrero, M. A.,
  Vazquez, R., \& Gomez, Y. 2001, MNRAS, 321, 487 

\reference{} Nulsen, P. E. J.,  David, L. P.,  McNamara, B. R.,  Jones, C.,
         Forman, W. R., \& Wise, M. 2002, ApJ, 568, 163 

\reference{} Omma, H., Binney, J. J., Bryan, G., \&
        Slyz, A. 2003, preprint (astro-ph/0307471)
        
\reference{} Quilis, V., Bower, R., \&  Balogh, M. L. 2001, MNRAS, 328, 1019
  
\reference{} Reynolds, C. S., Heinz, S., \& Begelman, M. C. 2001,
ApJ, 549,  L179



\reference{} Sabbadin, F., Turatto, M., Cappellaro, E.,
        Benetti, S., \& Ragazzoni, R., 2003, preprint 


\reference{} Sahai, R. 2000, ``Asymmetrical Planetary Nebulae II:
From Origins to Microstructures,'' eds. J.H. Kastner, N. Soker, \&
S. Rappaport, ASP Conf.\ Ser.\, Vol.\ 199, 209

\reference{} Sahai, R., \& Trauger, J. T. 1998, AJ, 116, 1357 

\reference{} Schindler, S., Castillo-Morales, A., De Filippis, E., Schwope, A.,
   \& Wambsganss, J. 2001, A\&A 376, L27

\reference{} Smith, D. A., Wilson, A. S., Arnaud, K. A., Terashima, Y.,
     \& Young, A. J. 2002, ApJ, 565, 195 

\reference{} Soker, N. 1994, AJ, 107, 276



\reference{} Soker, N. 2002, ApJ, 568, 726

\reference{} Soker, N. 2003, PASP, in press (astro-ph/0306616) (paper I)

\reference{} Soker, N., Blanton, E. L., \& Sarazin, C. L. 2002, ApJ, 573, 533
 
\reference{} Soker, N., \& Kastner, J. H. 2003, ApJ, 583, 368 
   


\reference{} Vrtilek, J. M., Grego, L., David, L. P.,
Ponman, T. J., Forman, W., Jones, C., \& Harris, D. E.  2002, APS, APRB, 17107
  
\reference{} White, R. E., III, \& Davis, D. S. 2003, in preparation.

\reference{} Zhekov, S. A., \& Myasnikov, A. V. 2000, ApJ, 543, L57 

\end{references}
\end{document}